\documentclass[pdflatex,sn-mathphys]{sn-jnl}

\jyear{2021}%

\usepackage{wasysym}
\usepackage{gensymb}
\usepackage{import}
\theoremstyle{thmstyleone}%
\theoremstyle{thmstyletwo}%

\theoremstyle{thmstylethree}%

\begin{document}

\title{Precise Calorimetry of Small Metal Samples Using Noise Thermometry}


\author*[1,2]{\fnm{Jan} \sur{Knapp}}\email{jan.knapp@rhul.ac.uk}

\author*[1]{\fnm{Lev V.} \sur{Levitin}}\email{l.v.levitin@rhul.ac.uk}

\author[1]{\fnm{J\'an} \sur{Ny\'eki}}

\author[2]{\fnm{Manuel} \sur{Brando}}

\author*[1]{\fnm{John} \sur{Saunders}}\email{j.saunders@rhul.ac.uk}

\affil[1]{\orgdiv{Department of Physics}, \orgname{Royal Holloway University of London}, \orgaddress{\street{Egham Hill}, \city{Egham}, \postcode{ TW20 0EX}, \country{UK}}}

\affil[2]{\orgdiv{Department of Physics}, \orgname{University of Oxford}, \orgaddress{\street{Parks Road}, \city{Oxford}, \postcode{ OX1 3PU}, \country{UK}}}

\affil[3]{\orgdiv{Physics of Quantum Materials}, \orgname{Max Planck Institute for Chemical Physics of Solids}, \orgaddress{\street{N\"othnitzer Stra\ss{}e 40}, \city{Dresden}, \postcode{01187}, \country{Germany}}}

\abstract{
We describe a compact calorimeter that opens ultra-low temperature heat capacity studies of small metal crystals in moderate magnetic fields.
The performance is demonstrated on the canonical heavy Fermion metal YbRh\textsubscript{2}Si\textsubscript{2}. Thermometry is provided by a fast current sensing noise thermometer.
This single thermometer enables us to cover a wide temperature range of interest from 175\,\textmu K to 90\,mK with temperature independent relative precision.
Temperatures are tied to the international temperature scale with a single point calibration.
A superconducting solenoid surrounding the cell provides the 
sample field for tuning its properties and operates a superconducting heat switch.
Both adiabatic and relaxation calorimetry techniques, as well as magnetic field sweeps, are employed.
The design of the calorimeter results in an addendum heat capacity which is negligible for the study reported.
The keys to sample and thermometer thermalisation are the lack of dissipation in the temperature measurement and the steps taken to reduce the parasitic heat leak into the cell to the tens of fW level.
}




\maketitle

\section{Introduction and Overview}

\begin{figure}[b]
    \centering
    \includegraphics[width=86mm]{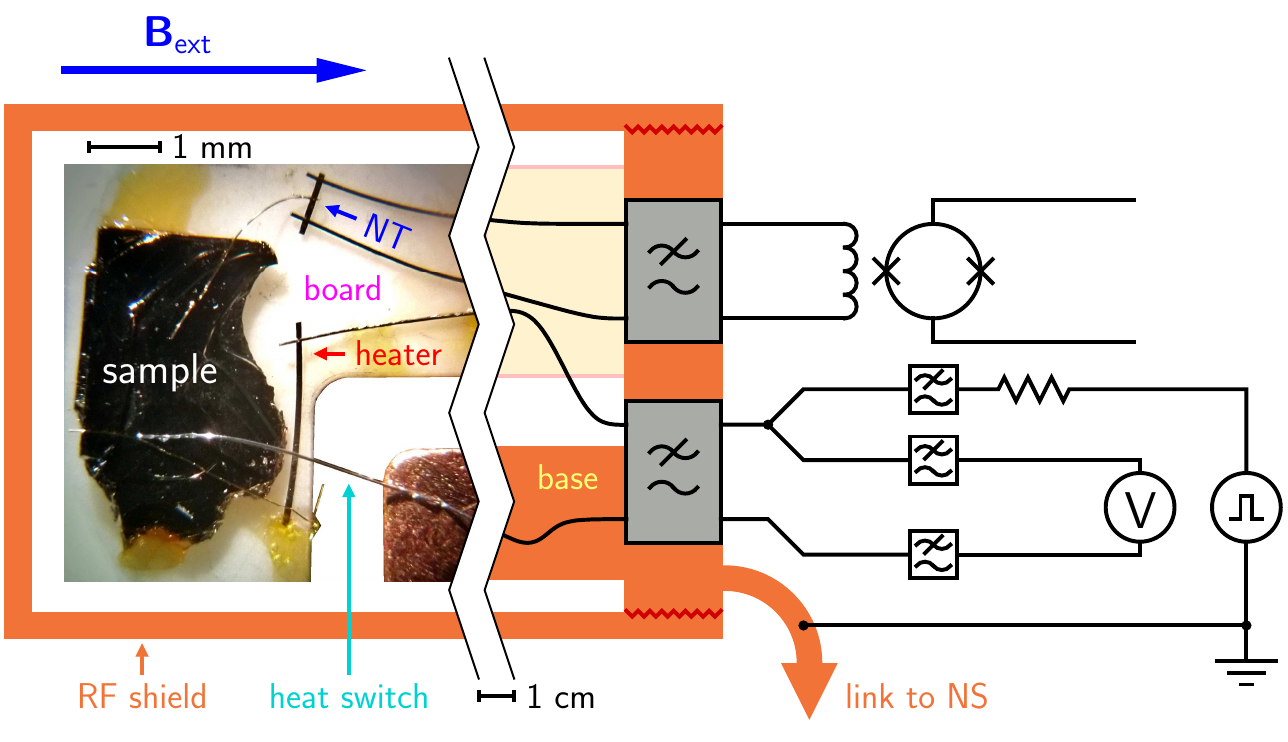}
    \caption{Design of the calorimetry cell.
    The key interior components are illustrated with a photograph, the rest of the setup is shown schematically.
    A sample (here a single crystal of YbRh$_2$Si$_2$)
    is thermalised to the copper base of the cell via an ultrasonically bonded aluminium heat switch.
    The copper base is cooled by the nuclear stage (NS) of an adiabatic demagnetisation refrigerator. Heater and noise thermometer (NT), made out of PtW ribbons, are connected to the sample via spot-welded gold wires, the electrical connections are made with spot-welded niobium wires. The sample, heater, and NT are glued to an alumina (sintered Al$_2$O$_3$) board with GE varnish. 
    The cell is radio-frequency (RF) tight with silver epoxy low-pass filters interrupting all electrical lines. The heater is driven by a pulse generator via a cold ballast resistor, with the heat switch and cryostat ground forming part of the current path.
    Additional electrical connections enable in situ measurement of the heater resistance. The cell is situated inside a compact superconducting solenoid, that provides the sample field $B_{\text{ext}}$ and controls the heat switch.}
    \label{fig:experiment}
\end{figure}

Measurements of thermodynamic quantities are key to understanding order and excitations in quantum materials.
For the measurement of heat capacity by the adiabatic method, the sample needs to be thermally separated from its environment, which poses technical challenges.
Numerous designs of calorimeters have been developed for the temperature range 0.01--1\,K, dealing with adiabatic isolation in a variety of ways \cite{Pobell2007,Brando2009,Wilhelm2004,Tsujii2005,Stewart1983,Bourgeois2005}, and references therein.
Surprisingly, ultra-low temperatures can simplify this issue because heat capacities and conductivities of structural materials can become vanishingly small.
However, complications in the choice of thermometry and the need to limit parasitic heat leaks into the experimental setup make heat capacity studies below 1\,mK relatively scarce.
Heat capacity measurements performed at ultra-low temperatures by the group in Bayreuth helped to resolve the interplay of (nuclear) magnetism and superconductivity in materials like pure aluminium \cite{Herrmannsdoerfer1998} and indium \cite{Herrmannsdoerfer2001} as well as the AuIn\textsubscript{2} alloy \cite{Herrmannsdoerfer1995,Herrmannsdoerfer1998}.
They achieved extremely low temperatures, often using the sample itself as the refrigerant.
However, such samples can be relatively large.

By contrast, quantum materials such as heavy fermion metals are often only available in the form of small single crystals.
Our calorimeter demonstrated outstanding performance with a state of the art 22\,mg single crystal sample of YbRh\textsubscript{2}Si\textsubscript{2} \cite{Krellner2012,Kliemt2019}, however, we argue that its simple design can be used for a wide range of materials.
Using pure and well characterised construction materials, with state of the art thermometry and shielding, we succeeded in measuring the heat capacity with unprecedented precision and accuracy over the temperature range 175\,\textmu K to 90\,mK \cite{Knapp2023}.

\begin{figure}[t]
    \centering
    \includegraphics{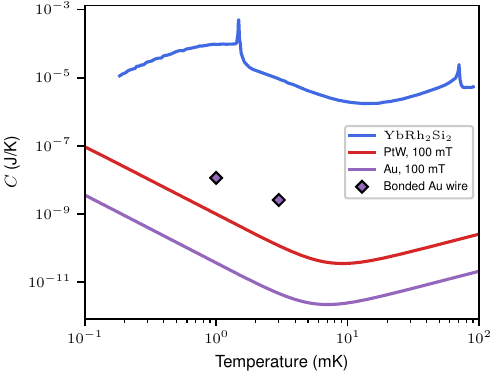}
    \caption{Measured heat capacity of YbRh\textsubscript{2}Si\textsubscript{2} vs. an estimate of the addendum heat capacity of the empty cell from the known amount of Au (6\,mm of $\diameter$25\,\textmu m) and PtW (3.5\,mm of $\diameter$50\,\textmu m) wires used, demonstrating that the addendum is negligible.
    The lines consist of a linear in $T$ electronic and a Schottky $T^{-2}$ nuclear heat capacity.
    Purple diamonds represent the heat capacity of a similar Au wire bonded to Au pads evaporated on silicon
    measured in zero magnetic field~\cite{Levitin2022a}. This relatively large value is attributed to the nuclear quadrupolar moments in crystallographically imperfect gold.
    All other structural materials of the cell either have a vanishingly small heat capacity, or the heat capacity is separated by a large enough thermal resistance, e.g. the nuclear heat capacity of aluminium inside Al\textsubscript{2}O\textsubscript{3}.
    }
    \label{fig:Addendum}
\end{figure}

The design of the calorimeter is illustrated in Fig.~\ref{fig:experiment}, as a combination of a microscope photograph and a schematic of the wiring.
All essential components in the cell are made and connected with the sample by small pieces of wire much smaller than the sample itself.
This was done by using well established techniques of spot-welding and ultrasonic wire bonding.
Such a compact design ensures the addendum heat capacity, which is the heat capacity of everything inside the cell apart from the sample, is negligible, as demonstrated in Fig.~\ref{fig:Addendum}.
This typical addendum serves as a reference for the selection of potential quantum material samples that might be studied with our technique.

Current sensing noise thermometry (CSNT) \cite{Levitin2023inprep,Shibahara2016,Casey2014,CASEY2003} was used over the full temperature range. Here it was implemented as a local probe on a small sample for calorimetry for the first time, with the sensor resistance selected for fast measurement time. The requirements are more stringent than in our recent use of noise thermometry to measure the heat capacity of a PrNi\textsubscript{5} nuclear demagnetisation stage \cite{Nyeki2022}.
The noise thermometer and a heater are both made out of Pt\textsubscript{92}W\textsubscript{8} (for simplicity PtW) ribbon, an alloy with negligible temperature dependence of resistivity, low magnetoresistance, and small heat capacity \cite{Ho1963,Ho1965}.
The ribbon was produced by flattening $\diameter$50\,\textmu m PtW wire.
The noise thermometer is located on the opposite side of the sample to the heater and an aluminium heat switch, helping to ensure that it accurately captures the temperature of the sample.
In the following, we describe the essential components in more detail, and report on their performance.

\section{Heat Switch}

To approach nearly perfect adiabatic isolation, the sample rests on a board made out of a good insulator, sintered Al\textsubscript{2}O\textsubscript{3} (alumina).
The 0.5\,mm thick board is fixed to the copper base, and extends 18\,mm into free space.
The base is well thermalised to our microkelvin cryostat via a cone joint.
The sample is precooled via a superconducting heat switch.
We have successfully used aluminium and lead.
In the former case, the sample is connected to the copper base by an ultrasonically wire bonded $\diameter$50\,\textmu m aluminium wire.
The aluminium heat switch operates at $\mu_0H_c=10$\,mT, and enabled measurements of  heat capacity by adiabatic methods in fields up to $H_c$.
The heat capacity can also be measured by relaxation methods in magnetic fields above $H_c$, albeit over a limited temperature range, governed by the normal state resistance of the heat switch.
The combination of the alumina board and aluminium heat switch results in excellent adiabatic isolation.
An upper bound on parasitic thermal conductance of 100\,fW/K was inferred from the undetectable change in the measured heat leak to the isolated sample, when sitting at a few hundred \textmu K, with the nuclear stage temperature deliberately raised to 10\,mK.
The normal state thermal conductance of the aluminium heat switch was sufficiently high to precool the sample well below 1\,mK in a couple of hours, and its operation at the end of the precool did not result in any observable heat release.

\begin{figure}[t]
    \centering
    \includegraphics[width=2.0in]{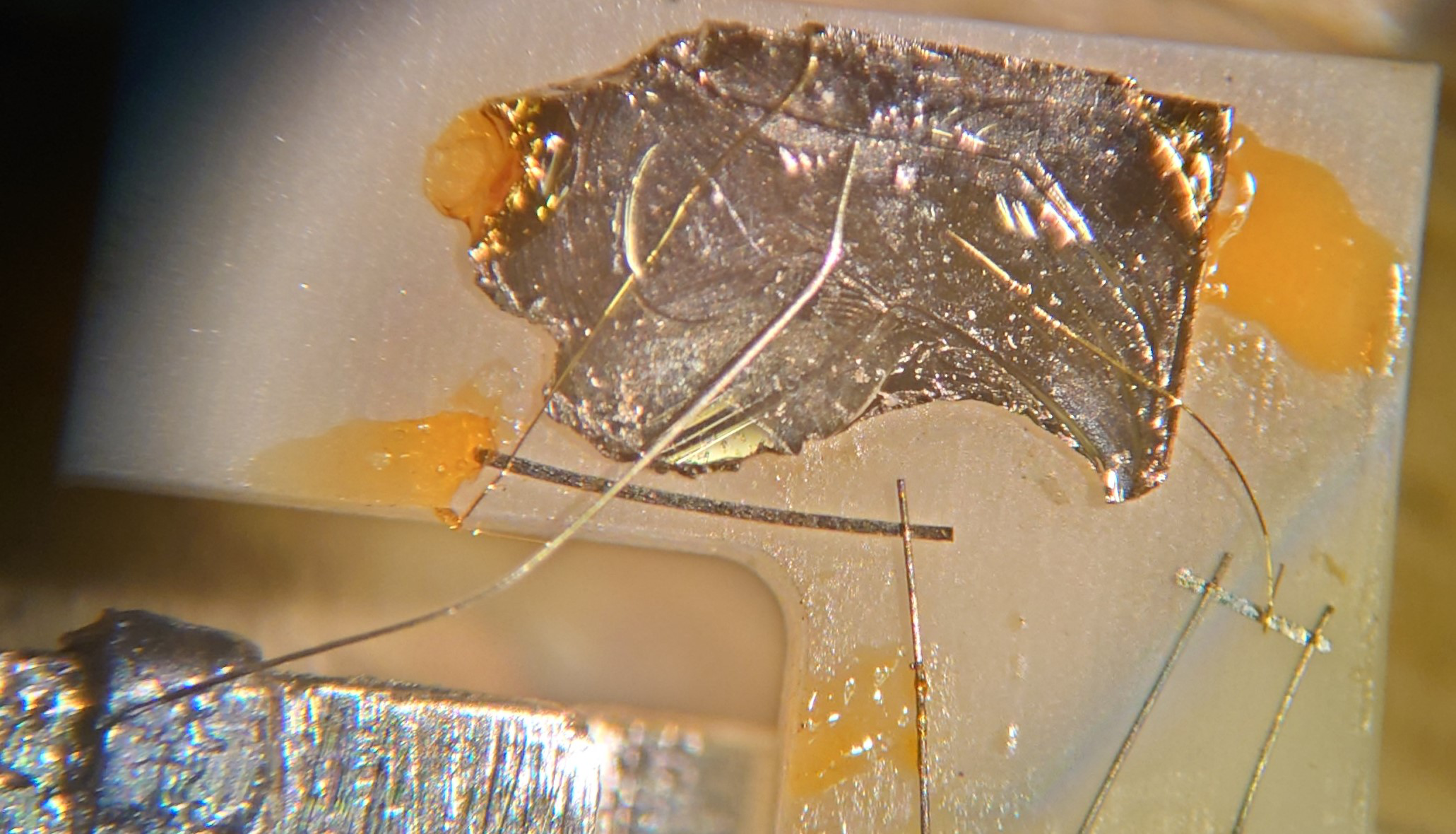}
    \caption{The calorimetry cell with the lead heat switch.
    The lead ``bead'' is visible in the bottom left corner, with the interconnecting silver wire spot-welded to the ``bead'' and the sample.}
    \label{fig:Pb_HS_cell}
\end{figure}

In order to extend the magnetic field range over which adiabatic methods can be employed, a second generation cell was constructed with a lead heat switch.
Lead is also a type I superconductor with a larger critical field of 80\,mT.
However, due to softness of lead it is difficult to obtain fine wires or use ultrasonic bonding.
The sample was moved from the copper base, shown in Fig.~\ref{fig:experiment}, to an identical base made out of pure silver, Fig.~\ref{fig:Pb_HS_cell}.
Lead is relatively easy to join with silver because the phase diagram of their mixtures features an eutectic point at $304\degree$C and 95.5\% atomic percentage of lead \cite{Karakaya1987}.
The eutectic character of the emerging Pb-Ag alloy ensures a clear transition from the superconductor to highly conducting normal metal across the joint.
We created a ``bead'' of pure lead on the silver cell base by melting it with a clean soldering iron on the surface of the base.
A pure $\diameter$50\,\textmu m silver wire was spot-welded to the YbRh\textsubscript{2}Si\textsubscript{2} sample and to the lead, making the construction of the heat switch relatively easy.

The ratio of normal and superconducting thermal conductivity, the switching ratio, depends on the Debye temperature $\theta$ of the material to the second power \cite{Pobell2007}.
Aluminium is therefore a highly favourable material for the construction of heat switches, because $\theta_{Al}=428$\,K,
while $\theta_{Pb}=105$\,K, making the expected switching ratio of an identical lead heat switch approximately 8 times smaller than aluminium.
We note that this is despite the much greater critical temperature of lead compared to aluminium.
The observed thermal leakage of our lead heat switch in the superconducting state was indeed significantly higher than its aluminium predecessor, becoming apparent above 4\,mK, instead of 10\,mK.
Nevertheless, measurements were still possible up to 90\,mK, thanks to the fast sampling of the noise thermometer.
Below 1\,mK, the thermal connection of sample to the sample base was again vanishingly small.
However, unlike with aluminium, the operation of the lead heat switch at the lowest temperature was always associated with a heat release, of the order of 10\,nJ.
Consequently, the starting temperature of the measurements
was typically 500\,\textmu K, even though the sample had been cooled down to 300\,\textmu K prior to opening the heat switch.
This heat release is much greater than the expected latent heat of the normal to superconducting transition of a type I superconductor
$
L(T)=-2\mu_0 H_c^2(T=0) (T / T_c)^2 \big[1-( T/T_c)^2 \big]V,
$
where $H_c$, $T_c$ and $V$ are the critical field, critical temperature and volume of the superconductor respectively~\cite{Poole2007}.
The latent heat calculated for the construction of our lead heat switch is in the fJ range, indicating some other source of heat release,
such as the dissipation when magnetic flux is expelled from the sample, since the lead bead has a rather bulky and irregular shape compared to the aluminium wire.

\section{Noise Thermometer}

Our calorimeter is distinguished by the use of current sensing noise thermometry (CSNT). 
The CSNT can be made exceptionally small, since the sensor itself is just a snippet of normal conductor.
In our case, a section of PtW ribbon with a relatively large resistance $R=203\,\mathrm{m\Omega}$ was selected.
This results in a large bandwidth over which the noise is measured, enabling fast temperature acquisition \cite{Levitin2023inprep,Shibahara2016,Casey2014,CASEY2003}.
This noise thermometer typically measures with a 1\% relative precision in about 50 seconds, independent of temperature, with the relative precision scaling as inverse square root of measuring time.

The thermal conductivity of the ribbon can be estimated using the Wiedemann-Franz law.
Such a large sensor resistance severely limits the parasitic heat leak to the thermometer that can be tolerated, to around 2\,fW in order to ensure thermal equilibrium within 10\% at 200\,$\mu$K.
This is achievable with appropriate filtering \cite{Levitin2022a}.
The CSNT method is essentially non-dissipative, requiring no external excitation.
A white noise voltage arises from thermal agitation of charge-carriers \cite{JOHNSON1927,Nyquist1928}.
The sample is connected to the input coil of a superconducting quantum interference device (SQUID) via a superconducting flux transformer.
A two stage SQUID sensor \cite{Drung2006,Drung2007} is used, operating as a sensitive current amplifier.
It is located remotely from the calorimeter, mounted on the still plate of the dilution refrigerator.
This technique is in principle a primary thermometer \cite{Shibahara2016,Kirste2023}, capable of yielding true thermodynamic temperature if all electrical parameters are known.
More conveniently, the thermometer can be calibrated with high accuracy at a single temperature, since the temperature dependence of the noise voltage derives from a fundamental physical law.
It is advantageous that the calibration point can be chosen at any temperature in the range of operation of the noise thermometer.
On our cryostat, a calibrated germanium resistance thermometer, compared to a \textsuperscript{3}He melting curve thermometer, was used to provide the calibration point for the cell CSNT at 200\,mK, where the calorimetry cell was in thermal equilibrium with both the nuclear stage and the mixing chamber.
This ties the measurement to the international temperature scale \cite{Kirste2023,Rusby2002,Mise2019}.

For measurements of heat capacity, fast temperature readings are essential.
This is achieved by measuring the current fluctuations over a large bandwidth.
The bandwidth $R / L_i$ of a CSNT is given by the resistance $R$ of the sensor and the SQUID input coil inductance $L_i=1.6\,$\textmu H, emphasizing the need for a high resistance sensor.
The thermometer bandwidth is typically still smaller than the full acquisition bandwidth capability of the SQUID amplifier.
In our case, $2^{17}$ points are recorded with 1\,MHz sampling frequency using a digital oscilloscope from National Instruments \cite{pxi5922}.
The recorded signal is Fourier transformed and the power spectral density averaged 200 times.
The flux noise power spectrum is
\begin{equation}
    S_\Phi = 4k_BTR\frac{M_i^2}{R^2+L_i^2\omega^2} + S_\Phi^0,
\label{eq:NoiseFunc}
\end{equation}
where $M_i=7.1$\,nH is the mutual inductance of the input coil and the first stage of the SQUID sensor, and $S_\Phi^0$ is the noise contribution to the measurement from the SQUID itself, which we consider to be white.
The comparison of the white noise produced by the sensor and the SQUID determines the noise temperature of the setup, which is 13\,\textmu K in our case, well below the lowest temperature ever measured.

\begin{figure}[t]
    \centering
    \includegraphics{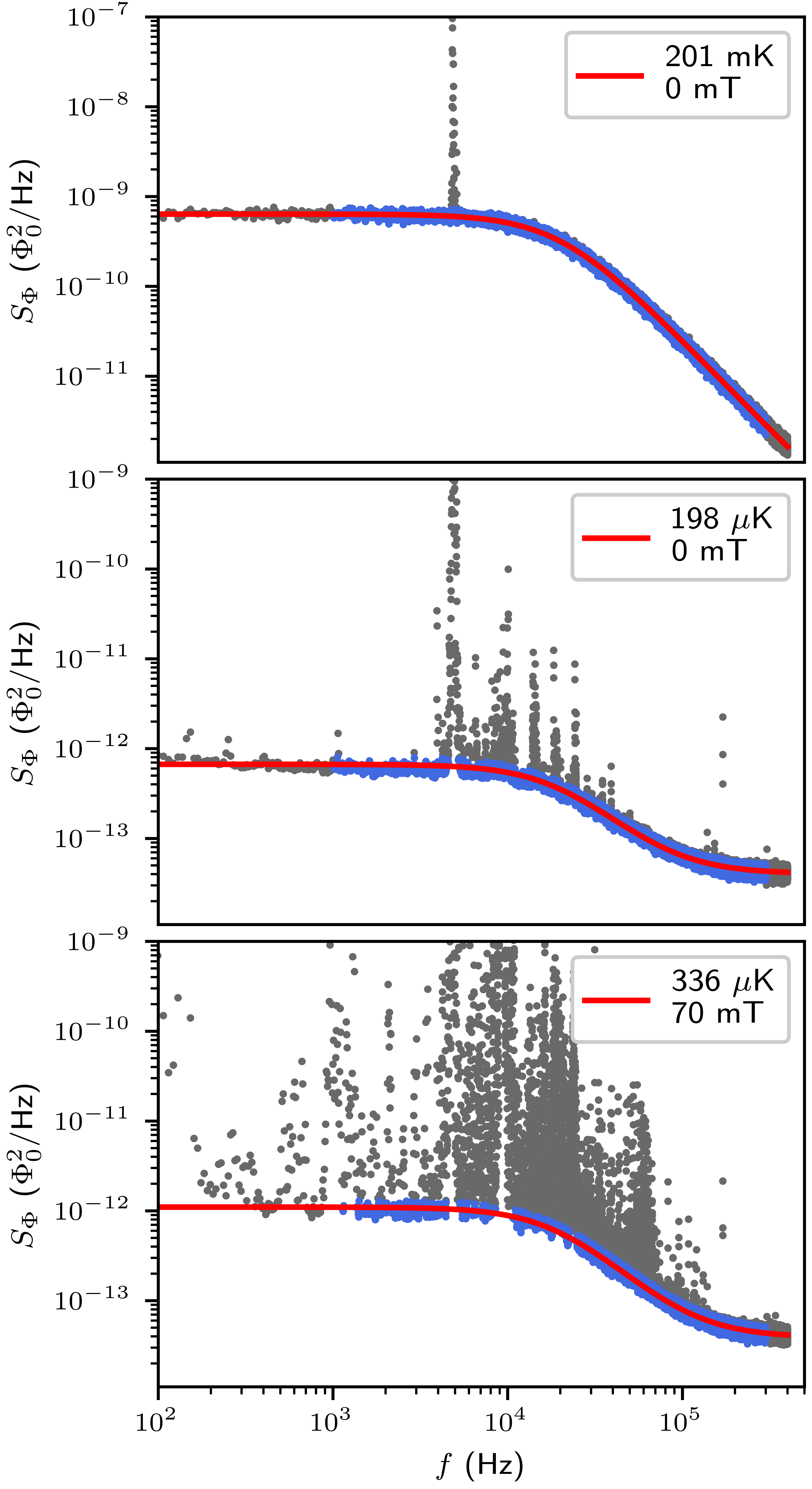}
    \caption{Noise thermometer spectra taken at three different temperatures and two applied fields.
    Points from the frequency range 1--300\,kHz (blue) are fitted with eq.~\ref{eq:NoiseFunc}, with $T$ and $S_{\Phi}^0$ as the fitting parameters.
    Grey points were discarded by an iterative discrimination procedure \cite{Levitin2023inprep}.
    In addition to the fitted terms, the $1/f$ contribution to the SQUID noise is apparent at low temperatures and frequencies.
    }
    \label{fig:Spectra_residuals}
\end{figure}

Fig.~\ref{fig:Spectra_residuals} shows examples of current sensing noise thermometer spectra taken at 198\,\textmu K and 201\,mK in zero magnetic field, accompanied by a spectrum taken at 336\,\textmu K in 70\,mT.
Whereas the 201\,mK spectrum is almost completely free from spurious noise, and can fitted by eq.~\ref{eq:NoiseFunc} after removing
a single peak at 4\,kHz, the spectra at ultra-low temperature, and especially in a magnetic field, are polluted by vibrational noise peaks to a much greater extent.
These peaks are easily removed by an iterative discrimination procedure, and do not significantly influence the measurement accuracy \cite{Levitin2023inprep}.
At the lowest temperatures the white noise $S_\Phi^0$, and a $1/f$ noise contributions of the SQUID become more significant. The SQUID white noise is determined from the highest frequency part of the spectrum. It is left as free parameter in the fits, since it can vary with the temperature of the SQUID itself, which is about 0.6\,K, the temperature of the still plate of our dilution refrigerator.
The $1/f$ noise has less influence in a large bandwidth noise thermometer, as the part of the spectrum below 1\,kHz is simply not used, as seen in Fig.~\ref{fig:Spectra_residuals}.
The maximum estimated uncertainty in $T$ after accounting for these last two contributions occurs at the lowest temperature, and is less than 5\%.

The resistive element of a CSNT can be made out of any normal metal, but the PtW alloy is advantageous in terms of resistivity and its negligible temperature dependence, low magnetoresistance, small heat capacity \cite{Ho1963,Ho1965}, and general metallurgical properties.
The PtW ribbon is spot-welded at both ends to $\diameter$50\,\textmu m Nb wires that act as thermal breaks.
The phase diagram of Pt and Nb features a rich number of alloys with very high melting points \cite{Tripathi1995}; we find that their spot-welds tend to be very stable and reliable both mechanically and resistively.
The noise thermometer is thermalised to the sample by a gold \diameter 25\,\textmu m wire, which is spot-welded to the middle of the PtW ribbon, and on the other end to the sample.
These spot-welds were also observed to be very reliable, despite a low solubility of Au and Pt \cite{Okamoto1985}.
Most importantly, they are not highly resistive, or superconducting, and do not deteriorate with time.

\section{Heater}

The ohmic heater with 0.9\,$\Omega$ resistance is also made out of the PtW ribbon.
The current enters via a spot-welded Nb wire serving as the $I_+$ lead.
On the other end of the ribbon, a spot-welded gold wire interconnects the heater with the sample.
The current flows through the relatively bulky sample, of negligible resistance, and into the cold cryostat ground ($I_-$) via the heat switch.
The heating pulse is produced by a voltage or current source at room temperature, connected between the $I_+$ current lead and the room temperature body of the cryostat, with a cold 100\,k$\Omega$ ballast resistor located on the current line at 0.6\,K protecting the cell from thermoelectric and other parasitic voltage sources.
Additional electrical connections to the heater enable in situ measurement of the heater resistance.

Figure~\ref{fig:experiment} shows that the cell is located inside of a continuous metal enclosure, formed by the copper body of the experimental platform and a silver epoxy filter and seal, developed originally to cool down two-dimensional electron gas to ultra-low temperatures \cite{Levitin2022a}.
The noise thermometer and heater ($I_+/V_+$ and $V_-$) leads were individually twisted and fully submerged into the silver epoxy \cite{epoxy} over a length of 82\,cm.
In our 4\,K tests, such silver epoxy low-pass filters were found to have a cut-off frequency of about 100\,MHz.
The heater wiring has an additional low-pass filter installed at the mixing chamber.
Inside the rf-shielded box of the filter, the $I_+$ and $V_+$ coaxial cables from room temperature were connected to the single ongoing $I_+/V_+$ lead, entering the cell.
Together with the $V_-$ lead, these three lines were each interrupted by 1\,k$\Omega$ + 10\,nF T-filters.
These filters, with approximately 15\,kHz cut-off frequency, further protected the heater from external interference.

\begin{figure}[t]
    \centering
    \includegraphics{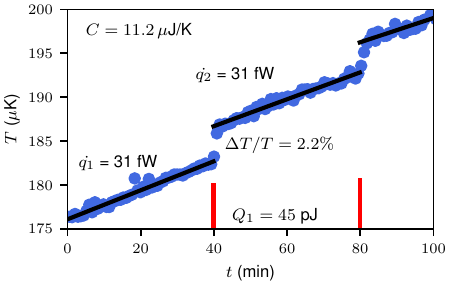}
    \caption{Typical heat pulses and temperature drifts at the lowest achieved temperature of 175\,\textmu K. Pulses (in red) of the order of 45\,pJ were applied every 40 minutes, resulting in units of percent increase of temperature, while the drifts in between them accurately determine the parasitic heat leak $\dot{q}_i$.
    }
    \label{fig:Adiabatic_pulse_method}
\end{figure}

\begin{figure}[t]
    \centering
    \includegraphics{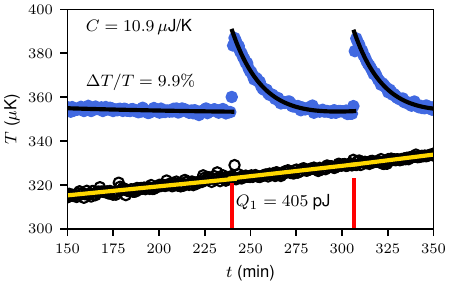}
    \caption{Example of the relaxation method in 60\,mT.
    The sample temperature (blues circles) shows relaxation from pulses, while the cryostat (black circles) is warming at the background.
    The cryostat's drifts between pulses are approximated by yellow lines.
    Internal time constants of the sample appear to be very small, the sample temperature rises to the full amplitude within two temperature readings, i.e. 100\,s.
    The relaxation after the pulse is fitted by a sum of exponential and linear functions, where the linear part uses the cryostat's drift as a guidance.}
    \label{fig:Relaxation_pulse_method}
\end{figure}

To apply heat, we used an  arbitrary waveform generator from National Instruments \cite{NIPXI5441}, or a Keithley 2400 SMU \cite{SMU}.
The former is most suited for producing short pulses, while the latter for producing constant heating current.
Extracting the heat capacity from heater pulses is the most common calorimetry method and can be done in the state of adiabatic isolation, as well as in the relaxation regime;
an example of the pulse measurement in the former regime, at the lowest achieved temperature of 175\,\textmu K, is shown in Fig.~\ref{fig:Adiabatic_pulse_method}, while typical relaxation from pulses, done in 60\,mT, is shown in Fig.~\ref{fig:Relaxation_pulse_method}.
The continuous warmup method, demonstrated in Fig.~\ref{fig:cont_warmup}, is a useful complement to the standard pulse method, especially in the vicinity of phase transitions.
Finally, magneto-caloric sweeps are introduced as another useful tool for characterising phase transitions and magnetic phases in Fig.~\ref{fig:MC_sweeps}.
In subsequent sections, we describe the results of these methods in more detail.

\section{Adiabatic Pulse Method}

Applying a pulse of energy $Q$ and observing the temperature increase from $T_l$ to $T_h$ as a result of the pulse, the heat capacity of the sample at temperature $T=(T_h+T_l)/2$ is
\begin{equation}
    C=\frac{Q}{\Delta T} ,
\end{equation}
where $\Delta T = T_h-T_l$.
We define the size of the pulse as $S_P=\Delta T/T$.
This value must be optimized based on technical performance of the cell.
As shown in Fig.~\ref{fig:Adiabatic_pulse_method}, the temperature drifts in-between pulses result from the parasitic heating $\dot{q}$ to the sample.
The longer the temperature drift is observed for, the more precise is the determination of $T_l$ and $T_h$.
The uncertainty in $T_{l,h}$ scales with the number of temperature readings in the drift $N$ as $1/\sqrt{N}$.
However, recording the drifts for a long time is not advantageous if the heat capacity is strongly temperature dependent, as in the vicinity  of a phase transition.
Thus it is essential to achieve a small parasitic heat leak into the calorimeter and a sufficiently high precision of the temperature, determined in a suitable measurement time (constrained by the temperature drift rate). Here these challenges are met by the large bandwidth noise thermometer, which achieves a 1\% precision in only 50 seconds, and the state of the art shielding of the cell, and filtering of the incoming leads, which reduces the parasitic heat leak to 30\,fW.
We believe that this heat leak is mostly deposited into the sample, since in a similar environment the heat leak
to the noise thermometer alone was found to be 0.1\,fW~\cite{Levitin2022a}.
Thanks to these technical achievements, it is possible to probe the sample with very small pulses, typically of the order of $S_P\approx1\%$, but also much less ($S_P\approx0.1\%$) where the heat capacity is large.
This allowed precise investigation of the low temperature phase transition $T_A$ in YbRh\textsubscript{2}Si\textsubscript{2} \cite{Knapp2023,Schuberth2016,Schuberth2022,Steppke2022} in the present application.

\begin{figure}[t]
    \centering
    \includegraphics{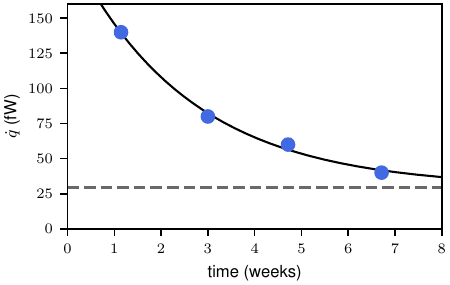}
    \caption{Time dependence of the heat leak into the YbRh\textsubscript{2}Si\textsubscript{2} calorimetry cell measured in zero magnetic field at temperatures around 1\,mK. This measurement reveals that there is an intrinsic, time dependent heat release, vanishing exponentially, and a residual heat leak of about 30\,fW.}
    \label{fig:Qdot_over_time}
\end{figure}

\begin{figure}[b]
    \centering
    \includegraphics{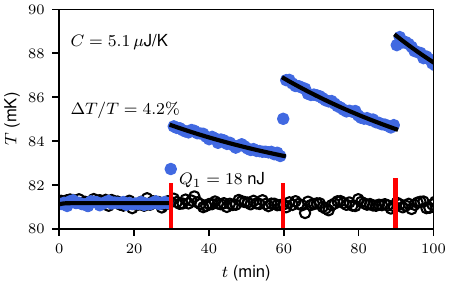}
    \caption{Typical heat pulses and temperature drifts towards the high end of the temperature spectrum.
    The temperature of the cryostat is stabilised and shown by black circles.
    }
    \label{fig:Pseudo_adiabatic_method}
\end{figure}

The parasitic heat leak is found to decay over time following initial cool-down, Fig.~\ref{fig:Qdot_over_time}. We attribute this to heat release of unknown origin in the sample itself or nearby components.
Nonetheless, the residual parasitic heat leak observed after about 8 weeks following the initial cooldown is very low.
Without application of external heat, the small sample would remain below 1\,mK for 20\,days, when initially cooled down to the lowest temperature of 175\,\textmu K, and it would take it additional 23\,days to warmup above the $T_A$ transition.

The adiabatic technique was also shown to work well at typical dilution refrigerator temperatures.
An example at 80\,mK is shown in Fig.~\ref{fig:Pseudo_adiabatic_method}.
The relaxation after the pulse arises from the thermal conductivity of the superconducting heat switch and the alumina board, so we refer to this method as pseudo-adiabatic.
Such measurements were possible up to 10\,mT for the aluminium switch and up to 65\,mT for the lead switch above which it leaked significantly.

\section{Relaxation Method}

The relaxation method was used in fields above 10\,mT, when the aluminium heat switch was in the normal state. 
An example of such a measurement done in 60\,mT is shown in Fig.~\ref{fig:Relaxation_pulse_method}.
It is possible to measure by this method if the internal time constants of the sample are much shorter than the time constant
$\tau\approx C/k$, where $C$ is the sample heat capacity and $k$ the thermal conductance of the heat switch in the normal state.
This requires the thermal diffusivity across the sample to be large and the coupling between various degrees of freedom of the sample
to be strong.
This is the case in this work.

In metals, the coupling of electrons and nuclei is given by the Korringa constant $\kappa=\tau_1T_e$, where $\tau_1$ is the nuclear spin-lattice relaxation time and $T_e$ the electronic temperature.
In metals with a strong hyperfine coupling of nuclei and electrons, as is the case of YbRh\textsubscript{2}Si\textsubscript{2}~\cite{Kondo1961,Bonville1984,Bonville1991,Nowik1968,Plessel2003,Knebel2006,Flouquet1975,Flouquet1978}, the Korringa constant is expected to be short, as established in PrNi\textsubscript{5} \cite{Pobell2007}. 
The immediate transfer of heat to the nuclear system apparent in figures~\ref{fig:Adiabatic_pulse_method} and~\ref{fig:Relaxation_pulse_method} supports this expectation of ``fast coupling''.

Following a pulse, thermal relaxation of the sample to a bath of constant temperature is typically exponential.
Incorporating the small temperature drifts, the fit function to the relaxation was $T(t)=a\exp(-t/\tau) + bt$, where $\tau$ is the thermal time constant.
In YbRh\textsubscript{2}Si\textsubscript{2}, it was possible to measure the heat capacity using the relaxation method up to about 4\,mK.
In this case, the limitations were the conductance of the heat switch increasing linearly with temperature, while the heat capacity of predominantly nuclear origin decreasing approximately as $T^{-1.5}$ in this temperature range.
This causes the thermal time constant $\tau$ to quickly become comparable to the noise thermometer acquisition time, which is where the method fails.
In our case, the normal state resistance of the aluminium heat switch was about 1.5\,m$\Omega$, dominated by the wire bond of Al to YbRh\textsubscript{2}Si\textsubscript{2}, not the wire itself.
This imperfection, caused presumably either by oxidation or just general unfitness of the wire bonding technique for the combination of these two materials, was quite fortunate in the end because, practically, the heat switch simultaneously provided fast precooling time of the sample and allowed relaxation measurements over a relatively broad temperature range.
Clearly, the combination of sample and thermal link can be optimised to extend the temperature range where the relaxation method can be employed, but this was beyond the scope of the present work.
As discussed earlier, measurements in fields greater than 10\,mT were performed with a lead heat switch.

\begin{figure}[t]
    \centering
    \includegraphics{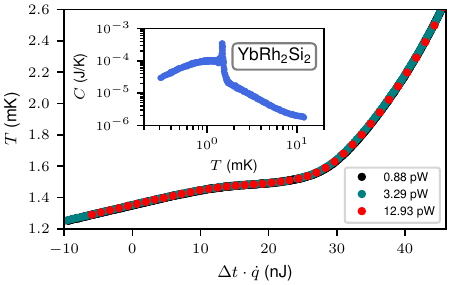}
    \caption{Continuous warmup across the $T_A$ low temperature phase transition in YbRh\textsubscript{2}Si\textsubscript{2} under three different heat loads. The warmup curves can be collapsed, scaling the time axis by the heating power $\dot{q}$. The inset shows heat capacity obtained from the warmup in 3.29\,pW, which manifests the peak at $T_A$ in its sharpness.
    The measurement also accurately agrees with the pulse method results, shown in Fig.~\ref{fig:Addendum}.
    }
    \label{fig:cont_warmup}
\end{figure}

\section{Continuous Warm-up Method}

The continuous warm-up method is extremely useful when studying anomalies and phase transitions.
The heat capacity is extracted from a continuous warmup record under the heat load $\dot{Q}$ as
\begin{equation}
    C=\frac{\dot{Q}}{\partial T/\partial t},
\end{equation}
where $\dot{Q}$ is the sum of the applied power and the parasitic heat leak; the effect of the latter is accounted for by repeated warm-ups under different power.
Here we probed the low temperature $T_A$ transition in YbRh\textsubscript{2}Si\textsubscript{2}, as shown in Fig.~\ref{fig:cont_warmup}.
Three different powers were applied continuously by the heater.
When the time axis is scaled by the heater power, a collapse of the warmup curves is achieved.
The inferred heat capacity is in perfect agreement with that inferred by other methods.
From this measurement, we infer that the $T_A$ transition is second order.
There is no signature of a plateau in the continuous warm-up curve in Fig.~\ref{fig:cont_warmup} that would indicated a latent heat.
Rather, we find a sharp $\lambda$-like anomaly.

\section{Field Sweeps}

\begin{figure}[t]
    \centering
    \includegraphics{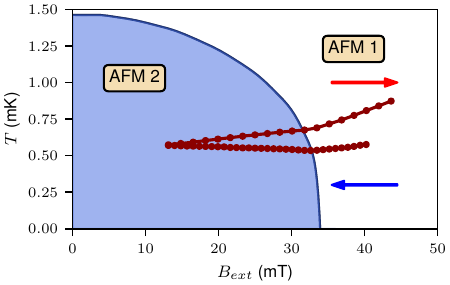}
    \caption{Magneto-caloric sweeps in YbRh\textsubscript{2}Si\textsubscript{2}.
    Blue and red coloured arrows indicate the direction of the sweeps as demagnetisation and magnetisation respectively.
    The vertical difference between the sweep up and down is proportional to the parasitic heat deposited in the sample during the sweep.
    The phase transition is clearly marked by a change of slope.
    }
    \label{fig:MC_sweeps}
\end{figure}

In this section, we describe the use of field sweeps in the detection of a field-induced phase transition in YbRh\textsubscript{2}Si\textsubscript{2}, exploiting the magneto-caloric effect.
In this context, YbRh\textsubscript{2}Si\textsubscript{2} is an example of a hyperfine enhanced nuclear refrigerant; its 4f electronic moment grows with the applied magnetic field, resulting in a much stronger field on the Yb nucleus, due the hyperfine interaction \cite{Knapp2023,Brando2013,Steinke2017}.
Our measurements of the moment growth (in the ab-crystallographic plane), combined with the hyperfine constant of Yb, 102\,T/$\mu_B$ \cite{Kondo1961,Bonville1984,Bonville1991}, determine the amplification ratio of the field applied to that felt by the nucleus to be about 150.
A magneto-caloric field sweeps of YbRh\textsubscript{2}Si\textsubscript{2} is shown in Fig.~\ref{fig:MC_sweeps}.
A clear feature in the field-induced temperature sweep is observed, marking a phase transition.
We observe the magneto-caloric effect to be stronger in the
primary antiferromagnetic order in YbRh\textsubscript{2}Si\textsubscript{2}
\cite{Knapp2023,Steppke2010,Krellner2009}, observed above about 35\,mT as $T\rightarrow 0$.
However, even in the low field phase, which we identified to be a form of a spatially modulated magnetic order \cite{Knapp2023}, the temperature can be lowered by demagnetisation, as was observed by a faster field sweep.
This technique has thus proved valuable to map the phase boundary near a quantum phase transition driven by magnetic field, where the critical temperature decreases strongly with field.

\section{Conclusions}

The design of the heat capacity cell described in this paper has the advantages of simplicity and modularity, which make it suitable for a large variety of quantum materials.
When adapting it for a particular metal sample, the outstanding question is always the common metallurgy of the sample and interconnecting wires and the addendum heat capacity, which determines which materials and bonding techniques to use.
The recent advances in fast current sensing noise thermometry \cite{Levitin2023inprep,Shibahara2016,Casey2014,CASEY2003} make this an attractive choice for calorimetry at dilution refrigerator temperatures and below.
Together with the filtering and shielding of superconducting leads we have implemented it brings calorimetry and bolometry (see review \cite{Giazotto2006}) to an improved level of sensitivity, in terms of sample size, and should have far-reaching consequences for future fundamental research.\\

\bmhead{Acknowledgments}
This work was supported by European Union’s Horizon 2020 Research and Innovation programme under Grant Agreement No. 824109 (European Microkelvin Platform) and the Deutsche Forschungsgemeinschaft (DFG, German Research Foundation) through Grants No. BR 4110/1-1, No. KR3831/4-1, and via the TRR 288, (422213477, project A03).
We thank Kristin Kliemt and Cornelius Krellner for providing the YbRh$_2$Si$_2$ samples, Vladimir Antonov for making the wire bonds to diverse materials, and Marijn Lucas for help with the magnet characterisation.
The cell was built and measurements were conducted at the London Low Temperature Laboratory, and we are grateful to Richard Elsom, Ian Higgs, Paul Bamford and Harpal Sandhu for excellent technical support.


\begin{thebibliography}{51}
\ifx \bisbn   \undefined \def \bisbn  #1{ISBN #1}\fi
\ifx \binits  \undefined \def \binits#1{#1}\fi
\ifx \bauthor  \undefined \def \bauthor#1{#1}\fi
\ifx \batitle  \undefined \def \batitle#1{#1}\fi
\ifx \bjtitle  \undefined \def \bjtitle#1{#1}\fi
\ifx \bvolume  \undefined \def \bvolume#1{\textbf{#1}}\fi
\ifx \byear  \undefined \def \byear#1{#1}\fi
\ifx \bissue  \undefined \def \bissue#1{#1}\fi
\ifx \bfpage  \undefined \def \bfpage#1{#1}\fi
\ifx \blpage  \undefined \def \blpage #1{#1}\fi
\ifx \burl  \undefined \def \burl#1{\textsf{#1}}\fi
\ifx \doiurl  \undefined \def \doiurl#1{\url{https://doi.org/#1}}\fi
\ifx \betal  \undefined \def \betal{\textit{et al.}}\fi
\ifx \binstitute  \undefined \def \binstitute#1{#1}\fi
\ifx \binstitutionaled  \undefined \def \binstitutionaled#1{#1}\fi
\ifx \bctitle  \undefined \def \bctitle#1{#1}\fi
\ifx \beditor  \undefined \def \beditor#1{#1}\fi
\ifx \bpublisher  \undefined \def \bpublisher#1{#1}\fi
\ifx \bbtitle  \undefined \def \bbtitle#1{#1}\fi
\ifx \bedition  \undefined \def \bedition#1{#1}\fi
\ifx \bseriesno  \undefined \def \bseriesno#1{#1}\fi
\ifx \blocation  \undefined \def \blocation#1{#1}\fi
\ifx \bsertitle  \undefined \def \bsertitle#1{#1}\fi
\ifx \bsnm \undefined \def \bsnm#1{#1}\fi
\ifx \bsuffix \undefined \def \bsuffix#1{#1}\fi
\ifx \bparticle \undefined \def \bparticle#1{#1}\fi
\ifx \barticle \undefined \def \barticle#1{#1}\fi
\bibcommenthead
\ifx \bconfdate \undefined \def \bconfdate #1{#1}\fi
\ifx \botherref \undefined \def \botherref #1{#1}\fi
\ifx \url \undefined \def \url#1{\textsf{#1}}\fi
\ifx \bchapter \undefined \def \bchapter#1{#1}\fi
\ifx \bbook \undefined \def \bbook#1{#1}\fi
\ifx \bcomment \undefined \def \bcomment#1{#1}\fi
\ifx \oauthor \undefined \def \oauthor#1{#1}\fi
\ifx \citeauthoryear \undefined \def \citeauthoryear#1{#1}\fi
\ifx \endbibitem  \undefined \def \endbibitem {}\fi
\ifx \bconflocation  \undefined \def \bconflocation#1{#1}\fi
\ifx \arxivurl  \undefined \def \arxivurl#1{\textsf{#1}}\fi
\csname PreBibitemsHook\endcsname

\bibitem{Pobell2007}
\begin{bbook}
\bauthor{\bsnm{Pobell}, \binits{F.}}:
\bbtitle{Matter and Methods at Low Temperatures}.
\bpublisher{Springer},
\blocation{Berlin, Heidelberg, New York}
(\byear{2007})
\end{bbook}
\endbibitem

\bibitem{Brando2009}
\begin{barticle}
\bauthor{\bsnm{Brando}, \binits{M.}}:
\batitle{Development of a relaxation calorimeter for temperatures between 0.05
  and 4 {K}}.
\bjtitle{Review of Scientific Instruments}
\bvolume{80}(\bissue{9}),
\bfpage{095112}
(\byear{2009}).
\doiurl{10.1063/1.3202380}
\end{barticle}
\endbibitem

\bibitem{Wilhelm2004}
\begin{barticle}
\bauthor{\bsnm{Wilhelm}, \binits{H.}},
\bauthor{\bsnm{Lühmann}, \binits{T.}},
\bauthor{\bsnm{Rus}, \binits{T.}},
\bauthor{\bsnm{Steglich}, \binits{F.}}:
\batitle{A compensated heat-pulse calorimeter for low temperatures}.
\bjtitle{Review of Scientific Instruments}
\bvolume{75}(\bissue{8}),
\bfpage{2700}--\blpage{2705}
(\byear{2004}).
\doiurl{10.1063/1.1771486}
\end{barticle}
\endbibitem

\bibitem{Tsujii2005}
\begin{barticle}
\bauthor{\bsnm{Tsujii}, \binits{H.}},
\bauthor{\bsnm{Andraka}, \binits{B.}},
\bauthor{\bsnm{Uchida}, \binits{M.}},
\bauthor{\bsnm{Tanaka}, \binits{H.}},
\bauthor{\bsnm{Takano}, \binits{Y.}}:
\batitle{Specific heat of the s=1 spin-dimer antiferromagnet
  {Ba$_3$Mn$_2$O$_8$} in high magnetic fields}.
\bjtitle{Physical Review B}
\bvolume{72}(\bissue{21}),
\bfpage{214434}
(\byear{2005}).
\doiurl{10.1103/physrevb.72.214434}
\end{barticle}
\endbibitem

\bibitem{Stewart1983}
\begin{barticle}
\bauthor{\bsnm{Stewart}, \binits{G.R.}}:
\batitle{Measurement of low{\textendash}temperature specific heat}.
\bjtitle{Review of Scientific Instruments}
\bvolume{54}(\bissue{1}),
\bfpage{1}--\blpage{11}
(\byear{1983}).
\doiurl{10.1063/1.1137207}
\end{barticle}
\endbibitem

\bibitem{Bourgeois2005}
\begin{barticle}
\bauthor{\bsnm{Bourgeois}, \binits{O.}},
\bauthor{\bsnm{Skipetrov}, \binits{S.E.}},
\bauthor{\bsnm{Ong}, \binits{F.}},
\bauthor{\bsnm{Chaussy}, \binits{J.}}:
\batitle{Attojoule calorimetry of mesoscopic superconducting loops}.
\bjtitle{Physical Review Letters}
\bvolume{94}(\bissue{5}),
\bfpage{057007}
(\byear{2005}).
\doiurl{10.1103/physrevlett.94.057007}
\end{barticle}
\endbibitem

\bibitem{Herrmannsdoerfer1998}
\begin{barticle}
\bauthor{\bsnm{Herrmannsdörfer}, \binits{T.}},
\bauthor{\bsnm{Rehmann}, \binits{S.}},
\bauthor{\bsnm{Seibold}, \binits{M.}},
\bauthor{\bsnm{Pobell}, \binits{F.}}:
\batitle{Interplay of nuclear magnetism and superconductivity}.
\bjtitle{Journal of Low Temperature Physics}
\bvolume{110}(\bissue{1/2}),
\bfpage{405}--\blpage{410}
(\byear{1998}).
\doiurl{10.1023/a:1022565607158}
\end{barticle}
\endbibitem

\bibitem{Herrmannsdoerfer2001}
\begin{barticle}
\bauthor{\bsnm{Herrmannsdörfer}, \binits{T.}},
\bauthor{\bsnm{Tayurskii}, \binits{D.}}:
\batitle{The impact of nuclear magnetism on superconductivity in a metal with
  nuclear electric quadrupole splitting: {Indium}}.
\bjtitle{Journal of Low Temperature Physics}
\bvolume{124}(\bissue{1/2}),
\bfpage{257}--\blpage{269}
(\byear{2001}).
\doiurl{10.1023/a:1017590221423}
\end{barticle}
\endbibitem

\bibitem{Herrmannsdoerfer1995}
\begin{barticle}
\bauthor{\bsnm{Herrmannsdörfer}, \binits{T.}},
\bauthor{\bsnm{Smeibidl}, \binits{P.}},
\bauthor{\bsnm{Schröder-Smeibidl}, \binits{B.}},
\bauthor{\bsnm{Pobell}, \binits{F.}}:
\batitle{Spontaneous nuclear ferromagnetic ordering of in nuclei in
  {AuIn}$_2$}.
\bjtitle{Physical Review Letters}
\bvolume{74}(\bissue{9}),
\bfpage{1665}--\blpage{1668}
(\byear{1995}).
\doiurl{10.1103/physrevlett.74.1665}
\end{barticle}
\endbibitem

\bibitem{Krellner2012}
\begin{barticle}
\bauthor{\bsnm{Krellner}, \binits{C.}},
\bauthor{\bsnm{Taube}, \binits{S.}},
\bauthor{\bsnm{Westerkamp}, \binits{T.}},
\bauthor{\bsnm{Hossain}, \binits{Z.}},
\bauthor{\bsnm{Geibel}, \binits{C.}}:
\batitle{Single-crystal growth of {YbRh$_2$Si$_2$} and {YbIr$_2$Si$_2$}}.
\bjtitle{Philosophical Magazine}
\bvolume{92}(\bissue{19-21}),
\bfpage{2508}--\blpage{2523}
(\byear{2012}).
\doiurl{10.1080/14786435.2012.669066}
\end{barticle}
\endbibitem

\bibitem{Kliemt2019}
\begin{barticle}
\bauthor{\bsnm{Kliemt}, \binits{K.}},
\bauthor{\bsnm{Peters}, \binits{M.}},
\bauthor{\bsnm{Feldmann}, \binits{F.}},
\bauthor{\bsnm{Kraiker}, \binits{A.}},
\bauthor{\bsnm{Tran}, \binits{D.-M.}},
\bauthor{\bsnm{Rongstock}, \binits{S.}},
\bauthor{\bsnm{Hellwig}, \binits{J.}},
\bauthor{\bsnm{Witt}, \binits{S.}},
\bauthor{\bsnm{Bolte}, \binits{M.}},
\bauthor{\bsnm{Krellner}, \binits{C.}}:
\batitle{Crystal growth of materials with the {ThCr}$_2${Si}$_2$ structure
  type}.
\bjtitle{Crystal Research and Technology}
\bvolume{55}(\bissue{2}),
\bfpage{1900116}
(\byear{2019}).
\doiurl{10.1002/crat.201900116}
\end{barticle}
\endbibitem

\bibitem{Knapp2023}
\begin{barticle}
\bauthor{\bsnm{Knapp}, \binits{J.}},
\bauthor{\bsnm{Levitin}, \binits{L.V.}},
\bauthor{\bsnm{Ny{\'{e}}ki}, \binits{J.}},
\bauthor{\bsnm{Ho}, \binits{A.F.}},
\bauthor{\bsnm{Cowan}, \binits{B.}},
\bauthor{\bsnm{Saunders}, \binits{J.}},
\bauthor{\bsnm{Brando}, \binits{M.}},
\bauthor{\bsnm{Geibel}, \binits{C.}},
\bauthor{\bsnm{Kliemt}, \binits{K.}},
\bauthor{\bsnm{Krellner}, \binits{C.}}:
\batitle{Electronuclear transition into a spatially modulated magnetic state in
  {YbRh$_2$Si$_2$}}.
\bjtitle{Physical Review Letters}
\bvolume{130}(\bissue{12}),
\bfpage{126802}
(\byear{2023}).
\doiurl{10.1103/physrevlett.130.126802}
\end{barticle}
\endbibitem

\bibitem{Levitin2022a}
\begin{barticle}
\bauthor{\bsnm{Levitin}, \binits{L.}},
\bauthor{\bparticle{van~der} \bsnm{Vliet}, \binits{H.}},
\bauthor{\bsnm{Theisen}, \binits{T.}},
\bauthor{\bsnm{Dimitriadis}, \binits{S.}},
\bauthor{\bsnm{Lucas}, \binits{M.}},
\bauthor{\bsnm{Corcoles}, \binits{A.}},
\bauthor{\bsnm{Ny{\'{e}}ki}, \binits{J.}},
\bauthor{\bsnm{Casey}, \binits{A.}},
\bauthor{\bsnm{Creeth}, \binits{G.}},
\bauthor{\bsnm{Farrer}, \binits{I.}},
\bauthor{\bsnm{Ritchie}, \binits{D.}},
\bauthor{\bsnm{Nicholls}, \binits{J.}},
\bauthor{\bsnm{Saunders}, \binits{J.}}:
\batitle{Cooling low-dimensional electron systems into the microkelvin regime}.
\bjtitle{Nature Communications}
\bvolume{13}(\bissue{1}),
\bfpage{1}--\blpage{8}
(\byear{2022}).
\doiurl{10.1038/s41467-022-28222-x}
\end{barticle}
\endbibitem

\bibitem{Levitin2023inprep}
\begin{botherref}
\oauthor{\bsnm{Levitin}, \binits{L.V.}},
\oauthor{\bparticle{et} \bsnm{al.}}:
Fast current sensing noise thermometry.
in preparation
(2024)
\end{botherref}
\endbibitem

\bibitem{Shibahara2016}
\begin{barticle}
\bauthor{\bsnm{Shibahara}, \binits{A.}},
\bauthor{\bsnm{Hahtela}, \binits{O.}},
\bauthor{\bsnm{Engert}, \binits{J.}},
\bauthor{\bparticle{van~der} \bsnm{Vliet}, \binits{H.}},
\bauthor{\bsnm{Levitin}, \binits{L.V.}},
\bauthor{\bsnm{Casey}, \binits{A.}},
\bauthor{\bsnm{Lusher}, \binits{C.P.}},
\bauthor{\bsnm{Saunders}, \binits{J.}},
\bauthor{\bsnm{Drung}, \binits{D.}},
\bauthor{\bsnm{Schurig}, \binits{T.}}:
\batitle{Primary current{\textendash}sensing noise thermometry in the
  millikelvin regime}.
\bjtitle{Philosophical Transactions of the Royal Society A: Mathematical,
  Physical and Engineering Sciences}
\bvolume{374}(\bissue{2064}),
\bfpage{20150054}
(\byear{2016}).
\doiurl{10.1098/rsta.2015.0054}
\end{barticle}
\endbibitem

\bibitem{Casey2014}
\begin{barticle}
\bauthor{\bsnm{Casey}, \binits{A.}},
\bauthor{\bsnm{Arnold}, \binits{F.}},
\bauthor{\bsnm{Levitin}, \binits{L.V.}},
\bauthor{\bsnm{Lusher}, \binits{C.P.}},
\bauthor{\bsnm{Ny{\'{e}}ki}, \binits{J.}},
\bauthor{\bsnm{Saunders}, \binits{J.}},
\bauthor{\bsnm{Shibahara}, \binits{A.}},
\bauthor{\bparticle{van~der} \bsnm{Vliet}, \binits{H.}},
\bauthor{\bsnm{Yager}, \binits{B.}},
\bauthor{\bsnm{Drung}, \binits{D.}},
\bauthor{\bsnm{Schurig}, \binits{T.}},
\bauthor{\bsnm{Batey}, \binits{G.}},
\bauthor{\bsnm{Cuthbert}, \binits{M.N.}},
\bauthor{\bsnm{Matthews}, \binits{A.J.}}:
\batitle{Current sensing noise thermometry: a fast practical solution to low
  temperature measurement}.
\bjtitle{Journal of Low Temperature Physics}
\bvolume{175},
\bfpage{764}--\blpage{775}
(\byear{2014}).
\doiurl{10.1007/s10909-014-1147-z}
\end{barticle}
\endbibitem

\bibitem{CASEY2003}
\begin{barticle}
\bauthor{\bsnm{Casey}, \binits{A.}}:
\batitle{Current{\textendash}sensing noise thermometry from 4.2 {K} to below 1
  {mK} using a {DC} {SQUID} preamplifier}.
\bjtitle{Physica B: Condensed Matter}
\bvolume{329-333},
\bfpage{1556}--\blpage{1559}
(\byear{2003}).
\doiurl{10.1016/s0921-4526(02)02293-7}
\end{barticle}
\endbibitem

\bibitem{Nyeki2022}
\begin{barticle}
\bauthor{\bsnm{Ny{\'{e}}ki}, \binits{J.}},
\bauthor{\bsnm{Lucas}, \binits{M.}},
\bauthor{\bsnm{Knappov{\'{a}}}, \binits{P.}},
\bauthor{\bsnm{Levitin}, \binits{L.V.}},
\bauthor{\bsnm{Casey}, \binits{A.}},
\bauthor{\bsnm{Saunders}, \binits{J.}},
\bauthor{\bparticle{van~der} \bsnm{Vliet}, \binits{H.}},
\bauthor{\bsnm{Matthews}, \binits{A.J.}}:
\batitle{High-performance cryogen-free platform for microkelvin-range
  refrigeration}.
\bjtitle{Physical Review Applied}
\bvolume{18}(\bissue{4}),
\bfpage{041002}
(\byear{2022}).
\doiurl{10.1103/physrevapplied.18.l041002}
\end{barticle}
\endbibitem

\bibitem{Ho1963}
\begin{barticle}
\bauthor{\bsnm{Ho}, \binits{J.C.}},
\bauthor{\bsnm{O’Neal}, \binits{H.R.}},
\bauthor{\bsnm{Phillips}, \binits{N.E.}}:
\batitle{Low temperature heat capacities of constantan and manganin}.
\bjtitle{Review of Scientific Instruments}
\bvolume{34}(\bissue{7}),
\bfpage{782}--\blpage{783}
(\byear{1963}).
\doiurl{10.1063/1.1718572}
\end{barticle}
\endbibitem

\bibitem{Ho1965}
\begin{barticle}
\bauthor{\bsnm{Ho}, \binits{J.C.}},
\bauthor{\bsnm{Phillips}, \binits{N.E.}}:
\batitle{Tungsten-platinum alloy for heater wire in calorimetry below
  0.1°{K}}.
\bjtitle{Review of Scientific Instruments}
\bvolume{36}(\bissue{9}),
\bfpage{1382}--\blpage{1382}
(\byear{1965}).
\doiurl{10.1063/1.1719917}
\end{barticle}
\endbibitem

\bibitem{Karakaya1987}
\begin{barticle}
\bauthor{\bsnm{Karakaya}, \binits{I.}},
\bauthor{\bsnm{Thompson}, \binits{W.T.}}:
\batitle{The {Ag-Pb} (silver-lead) system}.
\bjtitle{Bulletin of Alloy Phase Diagrams}
\bvolume{8}(\bissue{4}),
\bfpage{326}--\blpage{334}
(\byear{1987}).
\doiurl{10.1007/bf02869268}
\end{barticle}
\endbibitem

\bibitem{Poole2007}
\begin{bbook}
\bauthor{\bsnm{Poole}, \binits{C.P.}},
\bauthor{\bsnm{Farach}, \binits{H.A.}},
\bauthor{\bsnm{Creswick}, \binits{R.J.}},
\bauthor{\bsnm{Prozorov}, \binits{R.}}:
\bbtitle{Superconductivity},
p. \bfpage{650}.
\bpublisher{Elsevier/Academic Press},
\blocation{Amsterdam, Boston}
(\byear{2007})
\end{bbook}
\endbibitem

\bibitem{JOHNSON1927}
\begin{barticle}
\bauthor{\bsnm{Johnson}, \binits{J.B.}}:
\batitle{Thermal agitation of electricity in conductors}.
\bjtitle{Nature}
\bvolume{119}(\bissue{2984}),
\bfpage{50}--\blpage{51}
(\byear{1927}).
\doiurl{10.1038/119050c0}
\end{barticle}
\endbibitem

\bibitem{Nyquist1928}
\begin{barticle}
\bauthor{\bsnm{Nyquist}, \binits{H.}}:
\batitle{Thermal agitation of electric charge in conductors}.
\bjtitle{Physical Review}
\bvolume{32}(\bissue{1}),
\bfpage{110}--\blpage{113}
(\byear{1928}).
\doiurl{10.1103/physrev.32.110}
\end{barticle}
\endbibitem

\bibitem{Drung2006}
\begin{barticle}
\bauthor{\bsnm{Drung}, \binits{D.}},
\bauthor{\bsnm{Hinnrichs}, \binits{C.}},
\bauthor{\bsnm{Barthelmess}, \binits{H.{\textendash}.}}:
\batitle{Low{\textendash}noise ultra{\textendash}high{\textendash}speed dc
  {SQUID} readout electronics}.
\bjtitle{Superconductor Science and Technology}
\bvolume{19}(\bissue{5}),
\bfpage{235}--\blpage{241}
(\byear{2006}).
\doiurl{10.1088/0953-2048/19/5/s15}
\end{barticle}
\endbibitem

\bibitem{Drung2007}
\begin{barticle}
\bauthor{\bsnm{Drung}, \binits{D.}},
\bauthor{\bsnm{Assmann}, \binits{C.}},
\bauthor{\bsnm{Beyer}, \binits{J.}},
\bauthor{\bsnm{Kirste}, \binits{A.}},
\bauthor{\bsnm{Peters}, \binits{M.}},
\bauthor{\bsnm{Ruede}, \binits{F.}},
\bauthor{\bsnm{Schurig}, \binits{T.}}:
\batitle{Highly sensitive and easy{\textendash}to{\textendash}use {SQUID}
  sensors}.
\bjtitle{{IEEE} Transactions on Applied Superconductivity}
\bvolume{17}(\bissue{2}),
\bfpage{699}--\blpage{704}
(\byear{2007}).
\doiurl{10.1109/tasc.2007.897403}
\end{barticle}
\endbibitem

\bibitem{Kirste2023}
\begin{bchapter}
\bauthor{\bsnm{Kirste}, \binits{A.}},
\bauthor{\bsnm{Casey}, \binits{A.}},
\bauthor{\bsnm{Engert}, \binits{J.}},
\bauthor{\bsnm{Levitin}, \binits{L.}}:
\bctitle{Comparison of different johnson noise thermometers from millikelvin
  down to microkelvin temperatures}.
In: \bbtitle{Proceedings of the 29th International Conference on Low
  Temperature Physics (LT29)},
vol. \bseriesno{38}.
\bpublisher{Journal of the Physical Society of Japan},
\blocation{Sapporo, Japan}
(\byear{2023}).
\doiurl{10.7566/jpscp.38.011198}
\end{bchapter}
\endbibitem

\bibitem{Rusby2002}
\begin{barticle}
\bauthor{\bsnm{Rusby}, \binits{R.L.}},
\bauthor{\bsnm{Durieux}, \binits{M.}},
\bauthor{\bsnm{Reesink}, \binits{A.L.}},
\bauthor{\bsnm{Hudson}, \binits{R.P.}},
\bauthor{\bsnm{Schuster}, \binits{G.}},
\bauthor{\bsnm{Kühne}, \binits{M.}},
\bauthor{\bsnm{Fogle}, \binits{W.E.}},
\bauthor{\bsnm{Soulen}, \binits{R.J.}},
\bauthor{\bsnm{Adams}, \binits{E.D.}}:
\batitle{The provisional low temperature scale from 0.9 {mK to 1 K,
  PLTS-2000}}.
\bjtitle{Journal of Low Temperature Physics}
\bvolume{126}(\bissue{1/2}),
\bfpage{633}--\blpage{642}
(\byear{2002}).
\doiurl{10.1023/a:1013791823354}
\end{barticle}
\endbibitem

\bibitem{Mise2019}
\begin{botherref}
BIPM, SI Brochure – 9th Edition, Appendix 2, Mise en Pratique for the
  Definition of the Kelvin
(2019)
\end{botherref}
\endbibitem

\bibitem{pxi5922}
\begin{botherref}
24{\textendash}Bit, Flexible Resolution {PXI} Oscilloscope.
\url{http://www.ni.com/en-gb/support/model.pxi-5922.html}
\end{botherref}
\endbibitem

\bibitem{Tripathi1995}
\begin{barticle}
\bauthor{\bsnm{Tripathi}, \binits{S.N.}},
\bauthor{\bsnm{Bharadwaj}, \binits{S.R.}},
\bauthor{\bsnm{Dharwadkar}, \binits{S.R.}}:
\batitle{The {Nb-Pt} (niobium-platinum) system}.
\bjtitle{Journal of Phase Equilibria}
\bvolume{16}(\bissue{5}),
\bfpage{465}--\blpage{470}
(\byear{1995}).
\doiurl{10.1007/bf02645357}
\end{barticle}
\endbibitem

\bibitem{Okamoto1985}
\begin{barticle}
\bauthor{\bsnm{Okamoto}, \binits{H.}},
\bauthor{\bsnm{Massalski}, \binits{T.B.}}:
\batitle{The {Au-Pt} (gold-platinum) system}.
\bjtitle{Bulletin of Alloy Phase Diagrams}
\bvolume{6}(\bissue{1}),
\bfpage{46}--\blpage{56}
(\byear{1985}).
\doiurl{10.1007/bf02871187}
\end{barticle}
\endbibitem

\bibitem{epoxy}
\begin{botherref}
EPO-TEK E4110.
\url{https://www.epotek.com/docs/en/Datasheet/E4110.pdf}
\end{botherref}
\endbibitem

\bibitem{NIPXI5441}
\begin{botherref}
{43 MHz Bandwidth, 16\textendash Bit, Onboard Signal Processing PXI Waveform
  Generator}.
\url{https://www.ni.com/en-gb/support/model.pxi-5441.html}
\end{botherref}
\endbibitem

\bibitem{SMU}
\begin{botherref}
Keithley 2400.
\url{https://www.tek.com/en/products/keithley/source-measure-units/2400-standard-series-sourcemeter}
\end{botherref}
\endbibitem

\bibitem{Schuberth2016}
\begin{barticle}
\bauthor{\bsnm{Schuberth}, \binits{E.}},
\bauthor{\bsnm{Tippmann}, \binits{M.}},
\bauthor{\bsnm{Steinke}, \binits{L.}},
\bauthor{\bsnm{Lausberg}, \binits{S.}},
\bauthor{\bsnm{Steppke}, \binits{A.}},
\bauthor{\bsnm{Brando}, \binits{M.}},
\bauthor{\bsnm{Krellner}, \binits{C.}},
\bauthor{\bsnm{Geibel}, \binits{C.}},
\bauthor{\bsnm{Yu}, \binits{R.}},
\bauthor{\bsnm{Si}, \binits{Q.}},
\bauthor{\bsnm{Steglich}, \binits{F.}}:
\batitle{Emergence of superconductivity in the canonical
  heavy{\textendash}electron metal {YbRh}$_2${Si}$_2$}.
\bjtitle{Science}
\bvolume{351}(\bissue{6272}),
\bfpage{485}--\blpage{488}
(\byear{2016}).
\doiurl{10.1126/science.aaa9733}
\end{barticle}
\endbibitem

\bibitem{Schuberth2022}
\begin{botherref}
\oauthor{\bsnm{Schuberth}, \binits{E.}},
\oauthor{\bsnm{Wirth}, \binits{S.}},
\oauthor{\bsnm{Steglich}, \binits{F.}}:
Nuclear-order-induced quantum criticality and {Heavy-Fermion} superconductivity
  at ultra-low temperatures in {YbRh}$_2${Si}$_2$.
Frontiers in Electronic Materials
\textbf{2}
(2022).
\doiurl{10.3389/femat.2022.869495}
\end{botherref}
\endbibitem

\bibitem{Steppke2022}
\begin{barticle}
\bauthor{\bsnm{Steppke}, \binits{A.}},
\bauthor{\bsnm{Hamann}, \binits{S.}},
\bauthor{\bsnm{König}, \binits{M.}},
\bauthor{\bsnm{Mackenzie}, \binits{A.P.}},
\bauthor{\bsnm{Kliemt}, \binits{K.}},
\bauthor{\bsnm{Krellner}, \binits{C.}},
\bauthor{\bsnm{Kopp}, \binits{M.}},
\bauthor{\bsnm{Lonsky}, \binits{M.}},
\bauthor{\bsnm{Müller}, \binits{J.}},
\bauthor{\bsnm{Levitin}, \binits{L.V.}},
\bauthor{\bsnm{Saunders}, \binits{J.}},
\bauthor{\bsnm{Brando}, \binits{M.}}:
\batitle{Microstructuring {YbRh$_2$Si$_2$} for resistance and noise
  measurements down to ultra-low temperatures}.
\bjtitle{New Journal of Physics}
\bvolume{24}(\bissue{12}),
\bfpage{123033}
(\byear{2022}).
\doiurl{10.1088/1367-2630/aca8c6}
\end{barticle}
\endbibitem

\bibitem{Kondo1961}
\begin{barticle}
\bauthor{\bsnm{Kondo}, \binits{J.}}:
\batitle{Internal magnetic field in rare earth metals}.
\bjtitle{Journal of the Physical Society of Japan}
\bvolume{16}(\bissue{9}),
\bfpage{1690}--\blpage{1691}
(\byear{1961}).
\doiurl{10.1143/jpsj.16.1690}
\end{barticle}
\endbibitem

\bibitem{Bonville1984}
\begin{barticle}
\bauthor{\bsnm{Bonville}, \binits{P.}},
\bauthor{\bsnm{Imbert}, \binits{P.}},
\bauthor{\bsnm{J{\'{e}}hanno}, \binits{G.}},
\bauthor{\bsnm{Gonzalez{\textendash}Jimenez}, \binits{F.}},
\bauthor{\bsnm{Hartmann{\textendash}Boutron}, \binits{F.}}:
\batitle{Emission {M{\"{o}}ssbauer} spectroscopy and relaxation measurements in
  hyperfine levels out of thermal equilibrium:
  Very{\textendash}low{\textendash}temperature experiments on the {Kondo}
  {alloy Au$^{170}$Yb}}.
\bjtitle{Physical Review B}
\bvolume{30}(\bissue{7}),
\bfpage{3672}--\blpage{3690}
(\byear{1984}).
\doiurl{10.1103/physrevb.30.3672}
\end{barticle}
\endbibitem

\bibitem{Bonville1991}
\begin{barticle}
\bauthor{\bsnm{Bonville}, \binits{P.}},
\bauthor{\bsnm{Hodges}, \binits{J.A.}},
\bauthor{\bsnm{Imbert}, \binits{P.}},
\bauthor{\bsnm{J{\'{e}}hanno}, \binits{G.}},
\bauthor{\bsnm{Jaccard}, \binits{D.}},
\bauthor{\bsnm{Sierro}, \binits{J.}}:
\batitle{Magnetic ordering and paramagnetic relaxation of {Yb}$^{3+}$ in
  {YbNi}$_2${Si}$_2$}.
\bjtitle{Journal of Magnetism and Magnetic Materials}
\bvolume{97}(\bissue{1-3}),
\bfpage{178}--\blpage{186}
(\byear{1991}).
\doiurl{10.1016/0304-8853(91)90178-d}
\end{barticle}
\endbibitem

\bibitem{Nowik1968}
\begin{barticle}
\bauthor{\bsnm{Nowik}, \binits{I.}},
\bauthor{\bsnm{Ofer}, \binits{S.}}:
\batitle{{Mössbauer} studies of $^{170}${Yb} in several paramagnetic salts}.
\bjtitle{Journal of Physics and Chemistry of Solids}
\bvolume{29}(\bissue{12}),
\bfpage{2117}--\blpage{2119}
(\byear{1968}).
\doiurl{10.1016/0022-3697(68)90007-3}
\end{barticle}
\endbibitem

\bibitem{Plessel2003}
\begin{barticle}
\bauthor{\bsnm{Plessel}, \binits{J.}},
\bauthor{\bsnm{Abd{\textendash}Elmeguid}, \binits{M.M.}},
\bauthor{\bsnm{Sanchez}, \binits{J.P.}},
\bauthor{\bsnm{Knebel}, \binits{G.}},
\bauthor{\bsnm{Geibel}, \binits{C.}},
\bauthor{\bsnm{Trovarelli}, \binits{O.}},
\bauthor{\bsnm{Steglich}, \binits{F.}}:
\batitle{Unusual behavior of the low{\textendash}moment magnetic ground state
  of {YbRh}$_2${Si}$_2$ under high pressure}.
\bjtitle{Physical Review B}
\bvolume{67}(\bissue{18}),
\bfpage{180403}
(\byear{2003}).
\doiurl{10.1103/physrevb.67.180403}
\end{barticle}
\endbibitem

\bibitem{Knebel2006}
\begin{barticle}
\bauthor{\bsnm{Knebel}, \binits{G.}},
\bauthor{\bsnm{Boursier}, \binits{R.}},
\bauthor{\bsnm{Hassinger}, \binits{E.}},
\bauthor{\bsnm{Lapertot}, \binits{G.}},
\bauthor{\bsnm{Niklowitz}, \binits{P.G.}},
\bauthor{\bsnm{Pourret}, \binits{A.}},
\bauthor{\bsnm{Salce}, \binits{B.}},
\bauthor{\bsnm{Sanchez}, \binits{J.P.}},
\bauthor{\bsnm{Sheikin}, \binits{I.}},
\bauthor{\bsnm{Bonville}, \binits{P.}},
\bauthor{\bsnm{Harima}, \binits{H.}},
\bauthor{\bsnm{Flouquet}, \binits{J.}}:
\batitle{Localization of 4f state in {YbRh}$_2${Si}$_2$ under magnetic field
  and high pressure: Comparison with {CeRh}$_2${Si}$_2$}.
\bjtitle{Journal of the Physical Society of Japan}
\bvolume{75}(\bissue{11}),
\bfpage{114709}
(\byear{2006}).
\doiurl{10.1143/jpsj.75.114709}
\end{barticle}
\endbibitem

\bibitem{Flouquet1975}
\begin{barticle}
\bauthor{\bsnm{Flouquet}, \binits{J.}},
\bauthor{\bsnm{Brewer}, \binits{W.D.}}:
\batitle{Hyperfine interaction studies of local moments in metals}.
\bjtitle{Physica Scripta}
\bvolume{11}(\bissue{3-4}),
\bfpage{199}--\blpage{207}
(\byear{1975}).
\doiurl{10.1088/0031-8949/11/3-4/013}
\end{barticle}
\endbibitem

\bibitem{Flouquet1978}
\begin{barticle}
\bauthor{\bsnm{Flouquet}, \binits{J.}}:
\batitle{Kondo coupling, hyperfine and exchange interactions}.
\bjtitle{Le Journal de Physique Colloques}
\bvolume{39}(\bissue{C6}),
\bfpage{6}--\blpage{149361498}
(\byear{1978}).
\doiurl{10.1051/jphyscol:19786592}
\end{barticle}
\endbibitem

\bibitem{Brando2013}
\begin{barticle}
\bauthor{\bsnm{Brando}, \binits{M.}},
\bauthor{\bsnm{Pedrero}, \binits{L.}},
\bauthor{\bsnm{Westerkamp}, \binits{T.}},
\bauthor{\bsnm{Krellner}, \binits{C.}},
\bauthor{\bsnm{Gegenwart}, \binits{P.}},
\bauthor{\bsnm{Geibel}, \binits{C.}},
\bauthor{\bsnm{Steglich}, \binits{F.}}:
\batitle{Magnetization study of the energy scales in {YbRh}$_2${Si}$_2$ under
  chemical pressure}.
\bjtitle{Physica Status Solidi (B)}
\bvolume{250}(\bissue{3}),
\bfpage{485}--\blpage{490}
(\byear{2013}).
\doiurl{10.1002/pssb.201200771}
\end{barticle}
\endbibitem

\bibitem{Steinke2017}
\begin{barticle}
\bauthor{\bsnm{Steinke}, \binits{L.}},
\bauthor{\bsnm{Schuberth}, \binits{E.}},
\bauthor{\bsnm{Lausberg}, \binits{S.}},
\bauthor{\bsnm{Tippmann}, \binits{M.}},
\bauthor{\bsnm{Steppke}, \binits{A.}},
\bauthor{\bsnm{Krellner}, \binits{C.}},
\bauthor{\bsnm{Geibel}, \binits{C.}},
\bauthor{\bsnm{Steglich}, \binits{F.}},
\bauthor{\bsnm{Brando}, \binits{M.}}:
\batitle{Ultra{\textendash}low temperature ac susceptibility of the
  heavy{\textendash}fermion superconductor {YbRh$_2$Si$_2$}}.
\bjtitle{Journal of Physics: Conference Series}
\bvolume{807},
\bfpage{052007}
(\byear{2017}).
\doiurl{10.1088/1742-6596/807/5/052007}
\end{barticle}
\endbibitem

\bibitem{Steppke2010}
\begin{barticle}
\bauthor{\bsnm{Steppke}, \binits{A.}},
\bauthor{\bsnm{Brando}, \binits{M.}},
\bauthor{\bsnm{Oeschler}, \binits{N.}},
\bauthor{\bsnm{Krellner}, \binits{C.}},
\bauthor{\bsnm{Geibel}, \binits{C.}},
\bauthor{\bsnm{Steglich}, \binits{F.}}:
\batitle{Nuclear contribution to the specific heat of
  {Yb(Rh}$_{0.93}${Co}$_{0.07}$)$_2${Si}$_2$}.
\bjtitle{Physica Status Solidi (B)}
\bvolume{247}(\bissue{3}),
\bfpage{737}--\blpage{739}
(\byear{2010}).
\doiurl{10.1002/pssb.200983062}
\end{barticle}
\endbibitem

\bibitem{Krellner2009}
\begin{barticle}
\bauthor{\bsnm{Krellner}, \binits{C.}},
\bauthor{\bsnm{Hartmann}, \binits{S.}},
\bauthor{\bsnm{Pikul}, \binits{A.}},
\bauthor{\bsnm{Oeschler}, \binits{N.}},
\bauthor{\bsnm{Donath}, \binits{J.G.}},
\bauthor{\bsnm{Geibel}, \binits{C.}},
\bauthor{\bsnm{Steglich}, \binits{F.}},
\bauthor{\bsnm{Wosnitza}, \binits{J.}}:
\batitle{Violation of critical universality at the antiferromagnetic phase
  transition of {YbRh}$_2${Si}$_2$}.
\bjtitle{Physical Review Letters}
\bvolume{102}(\bissue{19}),
\bfpage{196402}
(\byear{2009}).
\doiurl{10.1103/physrevlett.102.196402}
\end{barticle}
\endbibitem

\bibitem{Giazotto2006}
\begin{barticle}
\bauthor{\bsnm{Giazotto}, \binits{F.}},
\bauthor{\bsnm{Heikkilä}, \binits{T.T.}},
\bauthor{\bsnm{Luukanen}, \binits{A.}},
\bauthor{\bsnm{Savin}, \binits{A.M.}},
\bauthor{\bsnm{Pekola}, \binits{J.P.}}:
\batitle{Opportunities for mesoscopics in thermometry and refrigeration:
  Physics and applications}.
\bjtitle{Reviews of Modern Physics}
\bvolume{78}(\bissue{1}),
\bfpage{217}--\blpage{274}
(\byear{2006}).
\doiurl{10.1103/revmodphys.78.217}
\end{barticle}
\endbibitem

\end{thebibliography}


\end{document}